\documentclass[journal]{IEEEtran}
\IEEEoverridecommandlockouts
\usepackage{cite}
\usepackage{breakurl}
\usepackage{amsmath,amssymb,amsfonts}
\usepackage{algorithmic}
\usepackage{graphicx}
\usepackage{textcomp}
\usepackage{xcolor}
\usepackage{subcaption}
\usepackage{empheq}
\def\BibTeX{{\rm B\kern-.05em{\sc i\kern-.025em b}\kern-.08em
    T\kern-.1667em\lower.7ex\hbox{E}\kern-.125emX}}

\usepackage{bbding}
\usepackage{pifont}
\usepackage{wasysym}
\usepackage{amssymb}

\usepackage[justification=centering]{caption}

\usepackage{diagbox}
\usepackage{mathtools}
\usepackage[left=0.65in,top=0.76in,right=0.65in,bottom=0.76in]{geometry}
\makeatletter
\let\old@ps@headings\ps@headings
\let\old@ps@IEEEtitlepagestyle\ps@IEEEtitlepagestyle
\def\confheader#1{
	\def\ps@IEEEtitlepagestyle{
		\old@ps@IEEEtitlepagestyle
		\def\@oddhead{\normalfont\footnotesize\centering#1}%
	}%
}

\makeatother

\confheader{%
	This paper has been accepted for publication in IEEE International Conference on Communications Workshops (ICC Workshops), 20-24 May 2019, Shanghai, China
}

\begin{document}

\title{Blockchain-based Lightweight Authentication Mechanism for Vehicular Fog Infrastructure}

\author{
	\IEEEauthorblockN{Kuljeet Kaur\IEEEauthorrefmark{1}, Member, IEEE, Sahil Garg\IEEEauthorrefmark{1}, Member, IEEE, Georges Kaddoum\IEEEauthorrefmark{1}, Member, IEEE, \\Fran\c{c}ois Gagnon\IEEEauthorrefmark{1}, Senior Member, IEEE, and Syed Hassan Ahmed\IEEEauthorrefmark{2}, Senior Member, IEEE}\\
	\IEEEauthorblockA{\IEEEauthorrefmark{1}Electrical Engineering Department, \'Ecole de technologie sup\'erieure, Montr\'eal, QC H3C 1K3, Canada.\\
		\IEEEauthorrefmark{2}Department of Computer Science, Georgia Southern University, Statesboro, GA 30460, USA.\\}
	(e-mail: kuljeet.kaur@ieee.org, sahil.garg@ieee.org, georges.kaddoum@etsmtl.ca, francois.gagnon@etsmtl.ca, \\ sh.ahmed@ieee.org)
}

\maketitle
\begin{abstract}
	With the increasing development of advanced communication technologies, vehicles are becoming smarter and more connected. Due to the tremendous growth of various vehicular applications, a huge amount of data is generated through advanced on-board devices and is deemed critical to improve driving safety and enhance vehicular services. However, cloud based models often fall short in applications where latency and mobility are critical. In order to fully realize the potential of vehicular networks, the challenges of efficient communication and computation need to be addressed. In this direction, vehicular fog computing (VFC) has emerged which extends the concept of fog computing to conventional vehicular networks. It is a geographically distributed paradigm that has the potential to conduct time-critical and data-intensive tasks by pushing intelligence (i.e. computing resources, storage, and application services) in the vicinity of end vehicles. However secure and reliable transmission are of significant importance in highly-mobile vehicular networks in order to ensure the optimal Quality of Service (QoS). In this direction, several authentication mechanisms have been proposed in the literature but most of them are found unfit due to absence of decentralization, anonymity, and trust characteristics. Thus, an effective cross-datacenter authentication and key-exchange scheme based on blockchain and elliptic curve cryptography (ECC) is proposed in this paper. Here, the distributed ledger of blockchain is used for maintaining the network information while the highly secure ECC is employed for mutual authentication between vehicles and road side units (RSUs). Additionally, the proposed scheme is lightweight and scalable for the considered VFC setup. The performance evaluation results against the existing state-of-the-art reveal that the proposed scheme accomplishes enhanced security features with reduced computational and communicational overheads. Further, its extensive evaluation on the widely applicable Automated Validation of Internet Security Protocols and Applications (AVISPA) tool guarantee its safeness against different attack vectors. 
\end{abstract}
	
	
\section{Introduction}
\begin{figure}
	\centering
	\fbox{\includegraphics[scale=0.4]{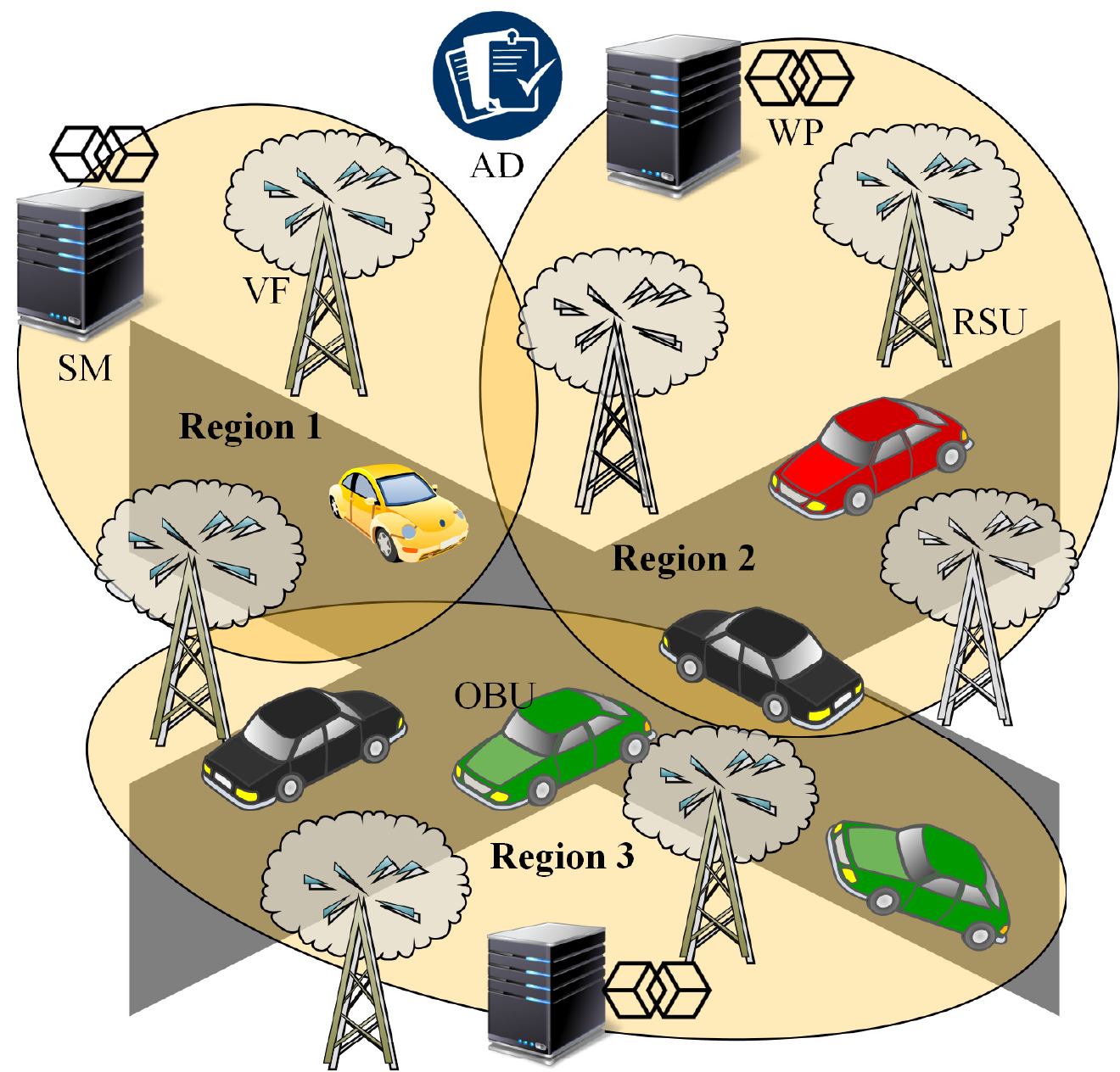}}
	\caption{A typical vehicular fog computing scenario \cite{8606185}.}
	\label{fig:SystemModel}
\end{figure}

With the emergence of information and communication technology, vehicles have become smarter than ever before. The integration of sensing, communication, and networking abilities in vehicles have enabled them to smartly interact with each other and with road-side units (RSU) in order to share information on a real-time basis. This information exchange capability of vehicles has led to the emergence of Vehicular adhoc Networks (VANETs), a key part of the intelligent transportation systems (ITS). These urban vehicular networks ensure a wide range of vehicle-based services for their intended users like road safety applications, smart traffic control, entertainment services, etc \cite{7724361}. Nevertheless, with the penetration of millions of smart vehicles in the global market, a gigantic bunch of vehicular on-board facilities have been furnished which collect and exchange huge amounts of data, resulting in significant growth of the network traffic \cite{8370877}. Although the conventional cloud computing architecture is known for providing a diverse range of services to vehicular networks, the long network distance between mobile devices and remote data centers hinders its performance while resulting in a high delay sensitivity \cite{8291113}. Thus, the concept of vehicular fog computing (VFC) has evolved, in which computation, storage, and networking are hosted in close proximity to vehicles \cite{7415983, 8493119}. The extension of the fog computing paradigm to conventional vehicular networks brings multiple opportunities to collect, process, organize, and store traffic data in real time. Locating the services close to end devices achieves better communication efficiency, seamless data processing, location awareness, and real-time response while minimizing latency along with other constraints. However, in order to ensure the desirable quality of service (QoS), an optimal balance between performance, security, and privacy requirements has become prominent \cite{huang2017vehicular, 8613868}.\\
\indent In this direction, several authentication protocols have been proposed for vehicular fog infrastructures. For example, Kang \textit{et al.} \cite{kang2018privacy} proposed a privacy-preserved pseudonym scheme to address the location privacy issues in VFC. Likewise, Yao \textit{et al.} \cite{yao2018reliable} presented a three-layered framework to ensure the reliability and security of VFC while preserving its performance. Considering the importance of vehicular crowdsensing in transportation, Wei \textit{et al.} \cite{wei2018privacy} proposed a fog-based privacy-preserving scheme for conditional road surface monitoring. In a similar direction, Basudan \textit{et al.} \cite{basudan2017privacy} also devised a privacy-preserving protocol to enhance the security of the vehicular crowdsensing network. To provide secure data transmission in VFC, Wang \textit{et al.} \cite{wang2018reliable} proposed a reliable and privacy-preserving task recomposition based on Paillier encryption. In \cite{8584096}, Kong \textit{et al.} also used homomorphic Paillier cryptosystem to design a secure querying scheme for data dissemination in VFC. Likewise, a new privacy-preserving route-sharing scheme for VFC was devised in \cite{li2018pros} to protect user and group privacy. Although several authentication schemes exists in the literature, most of them fail to provide data auditability, in case the central server crashes. Moreover, the existing approaches also incurs high computational cost and communication overhead due to large number of interactions among users and data centers \cite{kumar2016intelligent}. \\
\indent In order to provide information confidentiality, mutual authenticity, privacy, and anonymity, records should be kept in a distributed manner. Thus, blockchain has emerged as a feasible solution to achieve such goals. A blockchain is a decentralized and distributed ledger in which a digital record of transactions is maintained across blocks without requiring a central authority. All the blocks in the blockchain are linked in a chronological order to form a chain and secured using cryptography. Since, central managers are not involved in the blockchain structure, it is more robust against single point failure issues. Due to its wide range of benefits, blockchain-based applications, ranging from financial services to Internet of Things (IoT), are constantly expanding. Motivated by these facts, several approaches were developed in the literature in order to provide authentication in vehicular networks. For example, Lei \textit{et al.} \cite{lei2017blockchain} put forward a novel network topology based on a decentralized blockchain structure; wherein a blockchain was employed to simplify the distributed key management in vehicular communication systems. Kang \textit{et al.} \cite{8489897} adopted the concept of consortium blockchain to achieve secure data storage and sharing in vehicular edge networks. To support conditional privacy, one-to-many matching, destination matching, and data auditability in VFC, Li \textit{et al.} \cite{8452961} proposed an efficient and privacy preserving car-pooling scheme using blockchain. Accordingly, Yao \textit{et al.} \cite{8606185} presented a lightweight anonymous authentication mechanism for distributed vehicular fog services using blockchain. Their primary focus was on cross-data center authentication using blockchain and cryptographic functions. However, the designed scheme did not support mutual authentication between vehicles and service managers (SMs), which is a prerequisite for any secure authentication protocol. 
{
\subsection{Contributions} \label{sec:DesignGoals}
It is evident from the above discussion that a number of approaches have been proposed for VFC infrastructures but limited works have been carried out with respect to cross-data center authentication for accessing VFS, which is deemed important for providing any service in VFC scenarios. Keeping in view the necessity of accessing the VFS with reduced latency and overhead while maintaining an adequate level of security, it is essential to design a cross-datacenter authentication and key-exchange mechanism that is not only secure but also lightweight. Additionally, the literature also suggest that the existing approaches are primarily focused on centralized architectures which fall short in maintaining the network information across a distributed setup such as vehicular fogs in a profitable manner. Thus, the major contributions of the proposed work can be summarized as follows: 
\begin{itemize}
	\item We present an effective cross-datacenter authentication and key-exchange scheme using blockchain and Elliptic curve cryptography (ECC). Here, the distributed ledger of blockchain is used for maintaining the network information while the highly secure ECC is employed for mutual authentication between vehicles and RSUs. 
	\item The proposed scheme has been purposefully designed to be lightweight while provisioning the facility of re-authentication to participating vehicles. 
	\item The design goals of the proposed lightweight mechanism include cross-data center authentication, user anonymity, mutual authentication, user privacy and confidentiality.
	\item Lastly, the proposed scheme has been extensively evaluated with the existing state-of-the-art in terms of security features and computational and communicational overheads. Additionally, the safeness of the designed authentication and key-exchange mechanism has been validated using the well-known AVISPA tool. 
\end{itemize}

\subsection{Organization}
{The rest of this paper is organized as follows: Section~\ref{sec:SystemModel} illustrates the system model of the proposed work followed by the proposed authentication mechanism in Section~\ref{sec:ProposedScheme}. Moreover, detailed analysis of the security features and performance analysis are illustrated in Section~\ref{sec:ObservationAndAnalysis}. Finally, the conclusions are summarized in Section~\ref{sec:Conclusions}.}

\section{System Model} \label{sec:SystemModel}

A high level diagram considered in the proposed work is depicted in Fig.~\ref{fig:SystemModel} and is inspired from the work carried out in \cite{8606185, yao2018reliable}. As shown in the figure, the considered vehicular fog computing environment primarily consists of the following entities which actively participate in provisioning  the designed authentication and key exchange solution in the considered VFC setup based on blockchain and ECC. 

\vspace{1mm}
\noindent $\bullet$ \textbf{Region:} The considered VFC setup can be segregated into different regions comprised of different vehicular fog data-centers (VFDs) responsible of provisioning VFS to the intended users. Further each region may encapsulate one or many RSUs and a single service manager (SM) and witness peer (WP).  

\vspace{1mm}
\noindent $\bullet$ \textbf{Road Side Unit:} In VANETs, RSUs are the communicating nodes which provide different safety and entertainment services to vehicles and further relay the required information amongst each other. In the considered VFC scenario, a RSU manages a VFD and provides VFSs to the legitimate users. 

\vspace{1mm}
\noindent $\bullet$ \textbf{On-board Unit (OBU): }These are mounted on the vehicles and help them to interact with each other (V2V) and with RSUs (V2I). They are powered with communicational, computational, and storage facilities. In the current context, OBUs are referred to VFS end-users which are required to be registered with the audit department (AD) to access the intended services.

\vspace{1mm}
\noindent $\bullet$ \textbf{Audit Department:} Here, the AD is a central trusted authority which is responsible of publishing the public parameters to employed cryptographic functions. Additionally, it is also a place of registration for the vehicles/OBUs and the SMs.  

\vspace{1mm}
\noindent $\bullet$ \textbf{Service Manager:} The SM is essentially responsible for managing the blockchain network of a particular region. It is an authorized entity registered with the AD, which helps OBUs to authenticate and establish trust with the VFC infrastructure.

\vspace{1mm}
\noindent $\bullet$ \textbf{Witness Peer:} Every SM is associated with a witness peer (WP) that helps writing the authentication results to the public ledger. WP together with SM forms a consortium blockchain network and rely on Practical Byzantine Fault Tolerance (PBFT) for consensus establishment. 
}

\section{Cross Data Center Authentication in VFC} \label{sec:ProposedScheme}
The overall process of key exchange and authentication in the considered VFC environment can be segregated into the following five phases: 1) System Initialization Phase, 2) Registration Phase, 3) Mutual Authentication and Key Exchange Phase, 4) Consensus Phase, and 5) Service-Delivery Phase. The detailed description of these phases is presented below:

\vspace{3mm}
\noindent \textit{A. Phase I: System Initialization Phase}
\vspace{1mm}

During this phase, the AD prepares the VFC environment for the upcoming phases of the proposed scheme. The detailed elaboration of this phase is as follows:

\vspace{0.7mm}
\noindent \textit{Step 1:} The AD picks an elliptic curve $E$ with the public parameters $P$ and $n$. Here, the parameters refer to the base point of $E$ and a large prime number, respectively.

\vspace{0.7mm}
\noindent \textit{Step 2:} The AD deduces its public and private key as ($\mathbb{SK}_{AD}$ \& $\mathbb{PK}_{AD}$), where $\mathbb{SK}_{AD} \in Z_p^{*}$ and $\mathbb{PK}_{AD}= \mathbb{SK}_{AD}.P$. Here, the operation (.) denotes the ECC multiplicative operation. 

\vspace{0.7mm}
\noindent \textit{Step 3:} Finally, the AD declares the collision-resistant one-way hash functions ($H_1()$ and $H_2()$) to be used during the authentication and key exchange mechanism. With this, the parameters $<E, p, n, H_1(),H_2(), \mathbb{PK}_{AD}>$ are made public, i.e., are published.

\vspace{3mm}
\noindent \textit{B. Phase II:  Registration Phase}
\vspace{1mm}

\begin{figure}[t]
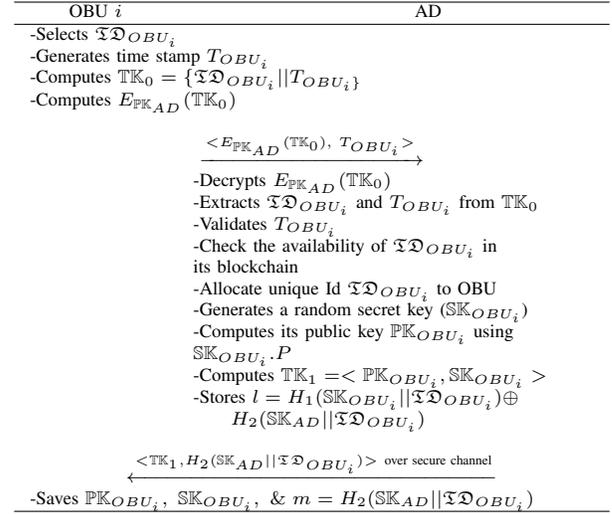

	\scriptsize
	\centering
	\begin{tabular}{p{1.75cm} p{.5cm} p{4.4cm}}
		\hline
		\multicolumn{1}{c}{{OBU $i$}} & {}  & \multicolumn{1}{c}{{AD}}   \\
		\hline
		\multicolumn{3}{l}{-Selects $\mathfrak{TD}_{OBU_i}$}\\
		\multicolumn{3}{l}{-Generates time stamp $T_{OBU_i}$}\\
		\multicolumn{3}{l}{-Computes $\mathbb{TK}_0=\{\mathfrak{TD}_{OBU_i} || T_{OBU_i\}}$}\\
		\multicolumn{3}{l}{-Computes $E_{\mathbb{PK}_{AD}}(\mathbb{TK}_0)$}\\
		&&\\
		\multicolumn{3}{c}{$\xrightarrow{<E_{\mathbb{PK}_{AD}}(\mathbb{TK}_0), ~T_{OBU_i}> }$}\\
		
		&\multicolumn{2}{l}{-Decrypts $E_{\mathbb{PK}_{AD}}(\mathbb{TK}_0)$}\\
		&\multicolumn{2}{l}{-Extracts  $\mathfrak{TD}_{OBU_i}$  and $T_{OBU_i}$ from $\mathbb{TK}_0$}\\
		&\multicolumn{2}{l}{-Validates $T_{OBU_i}$}\\
		& \multicolumn{2}{l}{-Check the availability of $\mathfrak{TD}_{OBU_i}$ in }\\
		& \multicolumn{2}{l}{its blockchain}\\
		&\multicolumn{2}{l}{-Allocate unique Id $\mathfrak{TD}_{OBU_i}$ to OBU}\\
		&\multicolumn{2}{l}{-Generates a random secret key ($\mathbb{SK}_{OBU_i}$)}\\
		&\multicolumn{2}{l}{-Computes its public key $ \mathbb{PK}_{OBU_i}$ using}\\
		& \multicolumn{2}{l}{$\mathbb{SK}_{OBU_i}.P$}\\
		& \multicolumn{2}{l}{-Computes $\mathbb{TK}_1=< \mathbb{PK}_{OBU_i}, \mathbb{SK}_{OBU_i}>$}\\
		&\multicolumn{2}{l}{-Stores $l =H_1(\mathbb{SK}_{OBU_i} || \mathfrak{TD}_{OBU_i}) \oplus$}\\
		&\multicolumn{2}{l}{ \hspace*{5mm}$H_2(\mathbb{SK}_{AD} || \mathfrak{TD}_{OBU_i}) $}\\
		&&\\
		\multicolumn{3}{c}{$\xleftarrow{<\mathbb{TK}_1, H_2(\mathbb{SK}_{AD} || \mathfrak{TD}_{OBU_i})>~\text{over secure channel}}$}\\
		\multicolumn{3}{l}{-Saves $\mathbb{PK}_{OBU_i},~\mathbb{SK}_{OBU_i},~\&~ m=H_2(\mathbb{SK}_{AD} || \mathfrak{TD}_{OBU_i})$}\\
		
		\hline
	\end{tabular}
	\caption{Phase II: Registration Phase.}
	\label{fig:Phase2}
\end{figure}

During this phase, the OBUs and SMs register themselves with the AD over a secure channel. In order to maintain the OBUs' and SMs' anonymity, their respective identities ($\mathfrak{TD}_{OBU_i}$ and $\mathfrak{TD}_{SM_j}$) are never relayed in clear text as illustrated in Fig.~\ref{fig:Phase2}.

\vspace{0.7mm}
\noindent \textit{Step 1:} The OBU $i$ selects an identity $\mathfrak{TD}_{OBU_i}$ for itself and simultaneously generates a time stamp $T_{OBU_i}$. This time stamp helps in validating the relayed message and ensures that the messages are not relayed in the near future.

\vspace{0.7mm}
\noindent \textit{Step 2:} The OBU then computes an intermediate token $\mathbb{TK}_0=\{\mathfrak{TD}_{OBU_i} || T_{OBU_i}\}$, followed by its encryption using the AD's public key $\mathbb{PK}_{AD}$. The computed token $\mathbb{TK}_0$ value is then relayed to AD. 

\vspace{0.7mm}
\noindent \textit{Step 3:} Upon receiving, $\mathbb{TK}_0$, the AD then decrypts the message to extract the values of $\mathfrak{TD}_{OBU_i}$ and $T_{OBU_i}$. Following this, AD validates the time-stamp of the generated token. If $T_{OBU_i}$ falls within the permissible time window, then AD proceeds, else the connection is terminated. In the next phase, the AD validates the availability of $\mathfrak{TD}_{OBU_i}$  in its blockchain. If a match is found, the $i^{th}$ OBU is directed to generate a new identity for itself and the above process is repeated. Afterwards, the AD generates the private-public key pairs ($\mathbb{PK}_{OBU_i}, \mathbb{SK}_{OBU_i}$) for the OBU and transmits the same over the secured channel. Additionally, the AD also computes, transmits, and stores the value of $l =H_1(\mathbb{SK}_{OBU_i} || \mathfrak{TD}_{OBU_i}) \oplus H_2(\mathbb{SK}_{AD} || \mathfrak{TD}_{OBU_i})$ in its distributed ledger.

\vspace{0.7mm}
\noindent \textit{Step 4:} The $i^{th}$ OBU then stores the values of $l$ and $ \mathbb{SK}_{OBU_i}$ in its repository.

\vspace{3mm}
\noindent \textit{C. Phase III: Mutual Authentication and Key Exchange Phase}\vspace{1mm}

\begin{figure}[ht]
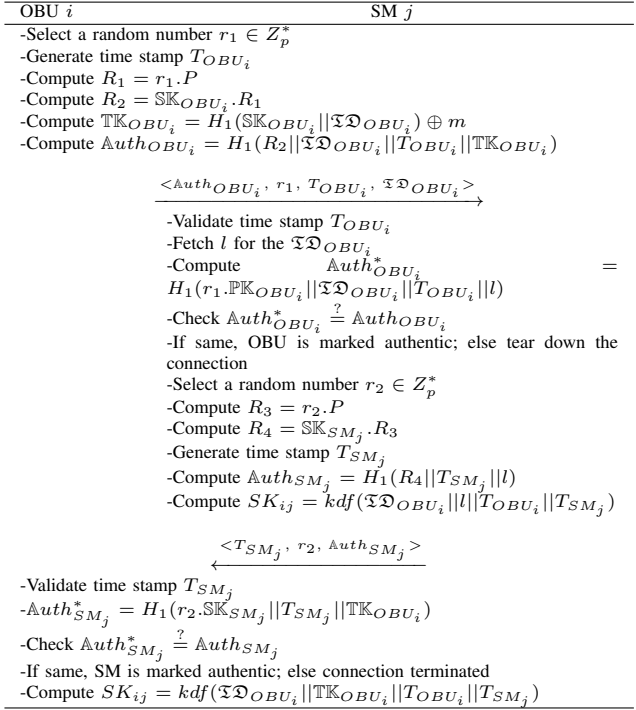

	\centering
	\scriptsize
	\begin{tabular}{p{.5cm} p{.4cm} p{6cm}}
		\hline
		\multicolumn{1}{c}{{OBU $i$}} & {}  & \multicolumn{1}{c}{{SM $j$}}   \\
		\hline
		\multicolumn{3}{l}{-Select a random number $r_1\in Z^{*}_p$ }\\
		\multicolumn{3}{l}{-Generate time stamp $T_{OBU_i}$ }\\
		\multicolumn{3}{l}{-Compute $R_1=r_1.P$ }\\
		\multicolumn{3}{l}{-Compute $R_2=\mathbb{SK}_{OBU_i}. R_1$ }\\
		\multicolumn{3}{l}{-Compute $\mathbb{TK}_{OBU_i} =  H_1(\mathbb{SK}_{OBU_i} || \mathfrak{TD}_{OBU_i}) \oplus m$}\\
		\multicolumn{3}{l}{-Compute $\mathbb{A}uth_{OBU_i}= H_1(R_2 || \mathfrak{TD}_{OBU_i} || T_{OBU_i} || \mathbb{TK}_{OBU_i}) $}\\
		&&\\
		\multicolumn{3}{c}{$\xrightarrow{<\mathbb{A}uth_{OBU_i},~r_1,~T_{OBU_i}, ~\mathfrak{TD}_{OBU_i} >}$}\\
		
		&& -Validate time stamp $T_{OBU_i}$\\
		&& -Fetch $l$ for the $\mathfrak{TD}_{OBU_i}$\\
		&& -Compute $\mathbb{A}uth_{OBU_i}^{*} = H_1(r_1. \mathbb{PK}_{OBU_i} || \mathfrak{TD}_{OBU_i} || T_{OBU_i} || l)  $\\
		&& -Check $\mathbb{A}uth_{OBU_i}^{*} \stackrel{?}{=}  \mathbb{A}uth_{OBU_i}$\\
		&& -If same, OBU is marked authentic; else tear down the connection\\
		&&-Select a random number $r_2\in Z^{*}_p$ \\
		&&-Compute $R_3=r_2.P$\\
		&&-Compute $R_4= \mathbb{SK}_{SM_j}.R_3$\\
		&& -Generate time stamp $T_{SM_j}$\\ 
		&& -Compute $\mathbb{A}uth_{SM_j} = H_1(R_4 || T_{SM_j} || l)$\\
		
		&& -Compute $SK_{ij} = kdf (\mathfrak{TD}_{OBU_i} || l || T_{OBU_i} || T_{SM_j} )$\\
		
		&&\\
		
		\multicolumn{3}{c}{$\xleftarrow{<T_{SM_j}, ~r_2, ~\mathbb{A}uth_{SM_j}>}$}\\
		
		\multicolumn{3}{l}{-Validate time stamp $T_{SM_j}$ }\\
		
		\multicolumn{3}{l}{-$\mathbb{A}uth_{SM_j}^{*} = H_1(r_2. \mathbb{SK}_{SM_j} || T_{SM_j}|| \mathbb{TK}_{OBU_i} )$ }\\
		\multicolumn{3}{l}{-Check $\mathbb{A}uth_{SM_j}^{*} \stackrel{?}{=}  \mathbb{A}uth_{SM_j}$ } \\
		\multicolumn{3}{l}{-If same, SM is marked authentic; else connection terminated }\\
		\multicolumn{3}{l}{-Compute $SK_{ij}= kdf (\mathfrak{TD}_{OBU_i} || \mathbb{TK}_{OBU_i} || T_{OBU_i} || T_{SM_j})$}\\
		\hline
		
	\end{tabular}
	\caption{Phase III: Mutual Authentication and Key Exchange Phase.}
	\label{fig:Phase3}
\end{figure}

During this phase, the OBUs mutually authenticate with the SMs via the deployed RSU and share a session key for further connection. The entire process is detailed in Fig.~\ref{fig:Phase3} and discussed as follows. 

\vspace{0.7mm}
\noindent \textit{Step 1:} The $i^{th}$ OBU initially generates a random number $r_1$ and time stamp $T_{OBU_i}$. Following this, it performs two ECC multiplicative operations over $r_1$ to compute $R_1$ and $R_2$.

\vspace{0.7mm}
\noindent \textit{Step 2:} Next, the OBU computes the values of the token $\mathbb{TK}_{OBU_i} =  H_1(\mathbb{SK}_{OBU_i} || \mathfrak{TD}_{OBU_i} ) \oplus m$; which involves a $H_1(.)$, concatenation and XOR operations. Using this token, the value of OBU's authentication token is estimated as follows: $\mathbb{A}uth_{OBU_i}= H_1(R_2 || \mathfrak{TD}_{OBU_i} || T_{OBU_i} || \mathbb{TK}_{OBU_i} )$. Finally, the values of $<\mathbb{A}uth_{OBU_i},~r_1,~T_{OBU_i}, ~\mathfrak{TD}_{OBU_i}>$ are relayed to the $j^{th}$ SM via its RSU for further analysis. 
	
\vspace{0.7mm}
\noindent \textit{Step 3:} Upon receiving these values, the first step performed by $SM_j$ is validation of the time-stamp $T_{OBU_i}$.  Upon successful validation, the SM fetches the value of $l$ from the registered list of OBU's blockchain. Using this value, SM calculates an authentication token to validate the truthfulness of $\mathbb{A}uth_{OBU_i}$ as follows: $\mathbb{A}uth_{OBU_i}^{*} = H_1(r_1. \mathbb{PK}_{OBU_i} || \mathfrak{TD}_{OBU_i} || T_{OBU_i} || l)$. If the values of $\mathbb{A}uth_{OBU_i}$ and $\mathbb{A}uth_{OBU_i}^{*}$ match, then SM successfully establishes the authenticity of the $i^{th}$ OBU, else the connection is terminated.
	
\vspace{0.7mm}
\noindent \textit{Step 4:} Next, $SM_j$ generates a random number, a time stamp ($T_{SM_j}$), and an authentication token ($\mathbb{A}uth_{SM_j}^{*} = H_1(r_2. \mathbb{SK}_{SM_j} || T_{SM_j}|| \mathbb{TK}_{OBU_i} )$) in the same manner as $OBU_i$. Additionally, it computes a session key, using key derivation function (kdf), to be used for further communication between the two parties as: $SK_{ij} = kdf (\mathfrak{TD}_{OBU_i} || l || T_{OBU_i} || T_{SM_j} )$. Finally, the values of $<T_{SM_j}, ~r_2, ~\mathbb{A}uth_{SM_j}>$ are transmitted to the $i^{th}$ OBU for further processing.
	
\vspace{0.7mm}
\noindent \textit{Step 5:} On receiving the values of $<T_{SM_j}, ~r_2, ~\mathbb{A}uth_{SM_j}>$, the OBU proceeds further only if the received time-stamp is found valid.

\vspace{0.7mm}
\noindent \textit{Step 6:} Next, the OBU validates the authenticity of the SM by performing the following computations: $\mathbb{A}uth_{SM_j}^{*} = H_1(r_2. \mathbb{SK}_{SM_j} || T_{SM_j}|| \mathbb{TK}_{OBU_i} )$. The successful validation of  $\mathbb{A}uth_{SM_j}^{*}$ and $\mathbb{A}uth_{SM_j}$ establishes the authenticity of the $j^{th}$ SM; this implies that both parties have successfully validated each other and are ready for further data transmission.

\vspace{0.7mm}
\noindent \textit{Step 7:} Finally, $OBU_i$ computes the session key $SK_{ij}= kdf (\mathfrak{TD}_{OBU_i} || \mathbb{TK}_{OBU_i} || T_{OBU_i} || T_{SM_j})$.

\vspace{3mm}
\noindent \textit{D. Phase IV: Consensus Phase}
\vspace{1mm}

In the proposed scheme, we consider a PFBFT consensus algorithm for forming the public ledger. The authentication results are transferred to the blockchain using the following steps. 

\vspace{0.7mm}
\noindent \textit{Step 1:} In the considered setup, we assume $n$ WPs with the ability to write a block to the public ledger. During the process of consensus, one of the WPs is marked as the ``Speaker"; while the rest act as ``Congressmen". The selected speaker is responsible for holding the consensus mechanism and cannot participate in the voting mechanism. However, the speaker is expected to conduct $m$ rounds of consensus for saving the time involved in speaker selection. Here, the speaker  $x$ is selected using the following evaluation: $x = (h\ mod\ n)+1$ where $h$ refers to the current block's height.

\vspace{0.7mm}
\noindent \textit{Step 2:} Post successful authentication and key exchange (as detailed in previous phase), the $j^{th}$ SM broadcasts  the authentication results to all the WPs.

\vspace{0.7mm}
\noindent \textit{Step 3:} Upon receiving the broadcast authentication results, the WPs store the results in their respective memories; before they can be transferred to the public ledger.

\vspace{0.7mm}
\noindent \textit{Step 4:} After $t$ intervals, the blockchain containing the authentication results is created, which in then followed by the voting process. In the initial run, the speaker broadcasts a request to the congressmen to vote using $< {P}_{req}, h, WP_x, block, Sig_{WP_x}(block)>$. Here, the variable ${P}_{req}$ denotes speaker's request others to vote.

\vspace{0.7mm}
\noindent \textit{Step 5:} Post receiving the request, the $k^{th}$ WP share its vote using $<{P}_{res}, h, WP_k, block, Sig_{WP_k}(block)>$ wherein ${P}_{res}$ denotes $k^{th}$ WP's response.

\vspace{0.7mm}
\noindent \textit{Step 6:} Upon receiving the response from the WPs, the speaker reaches a consensus to publish the block to the public ledger. 

\vspace{3mm}
\noindent \textit{E. Phase V: Service Delivery Phase}
\vspace{1mm}

This particular phase provides OBU the facility to avail the VFS seamlessly without the need to re-authenticate when moving to a new region. Under such a scenario, the $i^{th}$ OBU sends an encrypted token $E_{PK_{SM_{j^*}}}(l)$ where $l =H_1(\mathbb{SK}_{OBU_i} || \mathfrak{TD}_{OBU_i}) \oplus H_2(\mathbb{SK}_{AD} || \mathfrak{TD}_{OBU_i})$. Once the new SM, $j^*$, receives the request via the new RSU, then it decrypts the token to deduce $l$, and validates the existence of this token in its local database. If not found, it cross-checks the same in the public ledger. A match denotes that the authentication has been carried out sometime in the past. SM then checks the revocation list, if $l$ is not found then it is established that the $i^{th}$ OBU is valid and shall be seamlessly provided VFS without the need for re-authentication.

\section{Results and Discussion} \label{sec:ObservationAndAnalysis}
This section presents the performance evaluation of the proposed scheme against the current state-of-the-art. For the detailed performance evaluation, the comparison has been performed in terms of (1) security features (2) formal security verification using AVISPA and (3) computational and communicational overheads analysis. The corresponding details are mentioned as follows:

\vspace{2mm}
\noindent \textit{A. Security feature evaluation}
\vspace{1mm}

The design goals of the proposed scheme have been highlighted in Section~\ref{sec:DesignGoals}. In accordance with these features, the relative comparison with an existing scheme based on blockchain has been detailed in Table~\ref{tb:ComparisionAuthentication}. It is evident from the comparison that the proposed scheme provisions higher number of security features. For instance, the proposed scheme provides mutual authentication between OBUs and SMs, enables key exchange functionality, resists replay attacks, and supports forward secrecy. 

\begin{table}[ht]
	\centering
	\caption{Comparison of the  security features provided by proposed scheme against the existing state-of-the-art  \cite{8606185}.}
	\label{tb:ComparisionAuthentication}
	\scriptsize
	\begin{tabular}{p{4cm} p{1cm} p{1cm} }
		\hline
		{Scheme} & \cite{8606185}  & Proposed \\
		\hline
		\hline
		
		Supports Mutual Authentication & $\times$ &$\checkmark$ \\
		Supports Anonymity & $\checkmark$& $\checkmark$ \\
		Resists Replay &$\times$& $\checkmark$ \\
		
		Resists Impersonation &$\checkmark$&$\checkmark$ \\
		Supports Forward Secrecy &$\times$&$\checkmark$ \\
		Supports Confidentiality&$\checkmark$&$\checkmark$ \\
		Supports Integrity &$\checkmark$&$\checkmark$ \\
		Supports Non-interactivity &$\checkmark$&$\checkmark$ \\
		Supports Non-repudiation &$\checkmark$&$\checkmark$ \\
		Supports Key Exchange  & $\times$&$\checkmark$ \\
		\hline
		\\
	\end{tabular}	
\end{table}

\noindent \textit{B. Formal security verification}
\vspace{1mm}

In addition to the above validation, the execution of the proposed scheme was also formally verified using a well known tool-AVISPA. It is an open source suite of applications that is supportive in analyzing and verifying security protocols. In order to validate any designed security protocol, AVISPA needs the input in high level protocol specification language (HLPSL). The structure of HLPSL helps to describe the security protocol with intended security features and goals. The HLPSL defines the protocol in terms of different functions such as roles, transitions, composed role, and  a top-level role named environment. Further, AVISPA relies on the support of four different back-ends to validate any designed security protocol. These backends are namely-on the fly model checker (OFMC), CL-based attack searcher (CL-AtSe), SAT-based model checker (SATMC), and tree automata-based protocol analyzer (TA4SP). The back-ends help validate the safeness of the proposed mechanism against the targeted security goals and provide the user with a detailed trace in case of violation. \\
\indent In order to validate the proposed scheme, \textit{Phase III} of the scheme has been coded in HLPSL and subjected to AVISPA to verify its safeness against different security attacks. The related results, depicted in Fig.~\ref{fig:SPAN}, clearly show that the proposed authentication mechanism is safe on OFMC and CL-AtSe back-ends.

\begin{figure}[ht]
	\centering
	\begin{tabular}{| p{3.7cm} | p{3.7cm} |}
		\hline
		\begin{minipage}{15em}
			\scriptsize
			\ \ \\
			\% OFMC\\
			\% Version of 2006/02/13\\
			SUMMARY\\
			\hspace*{2mm}SAFE\\
			DETAILS\\
			\hspace*{2mm}BOUNDED\_NUMBER\\\hspace*{2mm}\_OF\_SESSIONS\\
			PROTOCOL\\
			\hspace*{2mm}/home/span/BlockECC.if\\
			GOAL\\
			\hspace*{2mm}as\_specified\\
			BACKEND\\
			\hspace*{2mm}OFMC\\
			COMMENTS\\
			STATISTICS\\
			\hspace*{2mm}parseTime: 0.00s\\
			\hspace*{2mm}searchTime: 0.23s\\
			\hspace*{2mm}visitedNodes: 27 nodes\\
			\hspace*{2mm}depth: 4 plies
		\end{minipage} & 
		\begin{minipage}{15em}
			\scriptsize
			\ \ \\
			SUMMARY\\
			\hspace*{2mm}SAFE\\
			DETAILS\\
			\hspace*{2mm}BOUNDED\_NUMBER\\hspace*{2mm}\_OF\_SESSIONS\\
			TYPED\_MODEL\\
			PROTOCOL\\
			\hspace*{2mm}/home/span/BlockECC.if\\
			GOAL\\
			\hspace*{2mm}As Specified\\
			BACKEND\\
			\hspace*{2mm}CL-AtSe\\
			STATISTICS\\
			\hspace*{2mm}Analysed   : 0 states\\
			\hspace*{2mm}Reachable  : 0 states\\
			\hspace*{2mm}Translation: 0.13 seconds\\
			\hspace*{2mm}Computation: 0.00 seconds
		\end{minipage}\\
		\hline
	\end{tabular}
	\caption{Evaluation of mutual authentication and key exchange mechanism on AVISPA}\label{fig:SPAN}
\end{figure}

\vspace{2mm}
\noindent \textit{C. Computational and Communicational Overhead Analysis}
\vspace{1mm}


\begin{figure*}[ht]
	\begin{subfigure}[t]{0.23\textwidth}
		\centering
		\includegraphics[scale=.33]{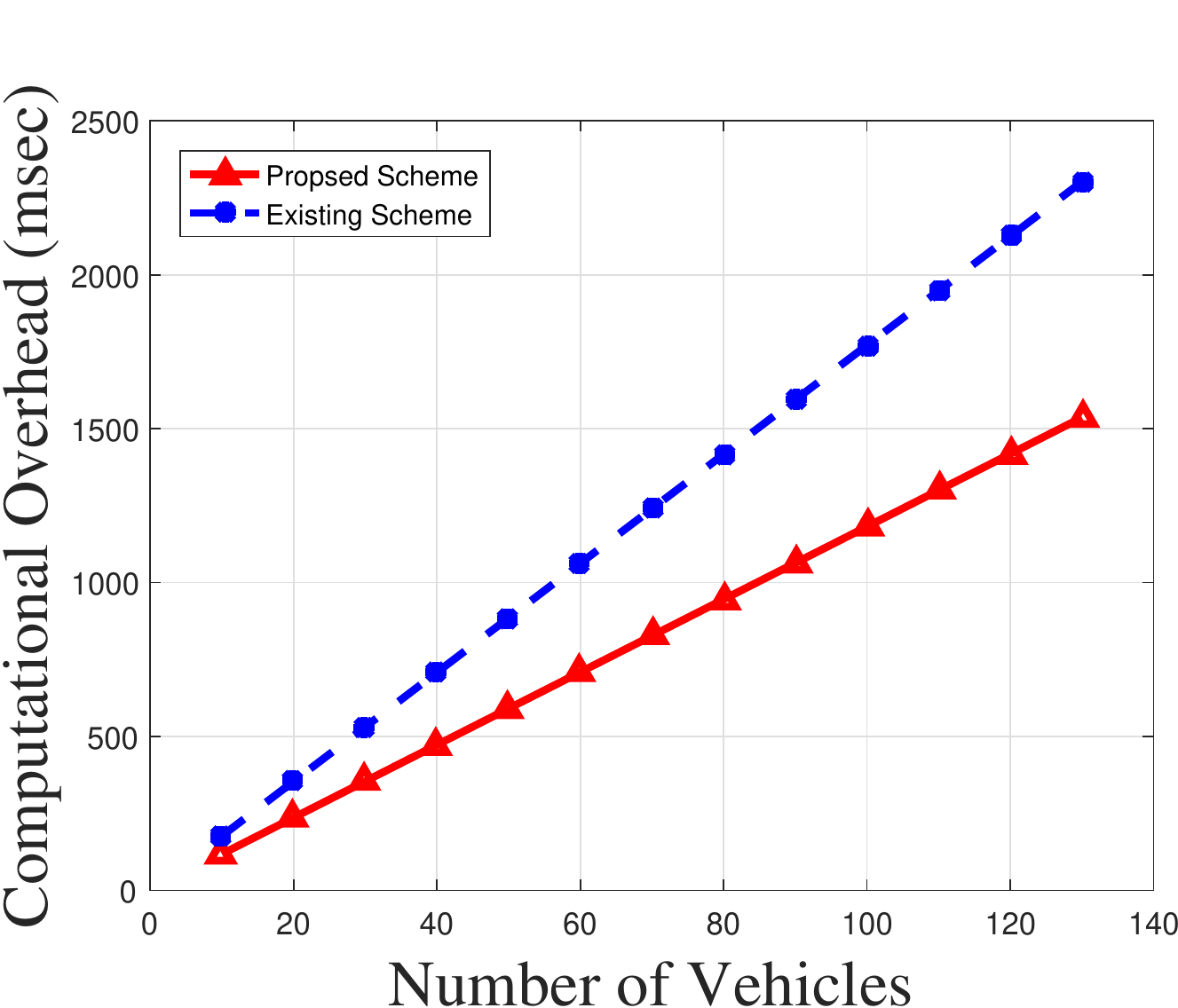}
		\caption{Computational overhead Vs number of vehicles.}
		\label{fig:computationaloverheadv}
	\end{subfigure}
	~
	\begin{subfigure}[t]{0.23\textwidth}
		\centering
		\includegraphics[scale=.33]{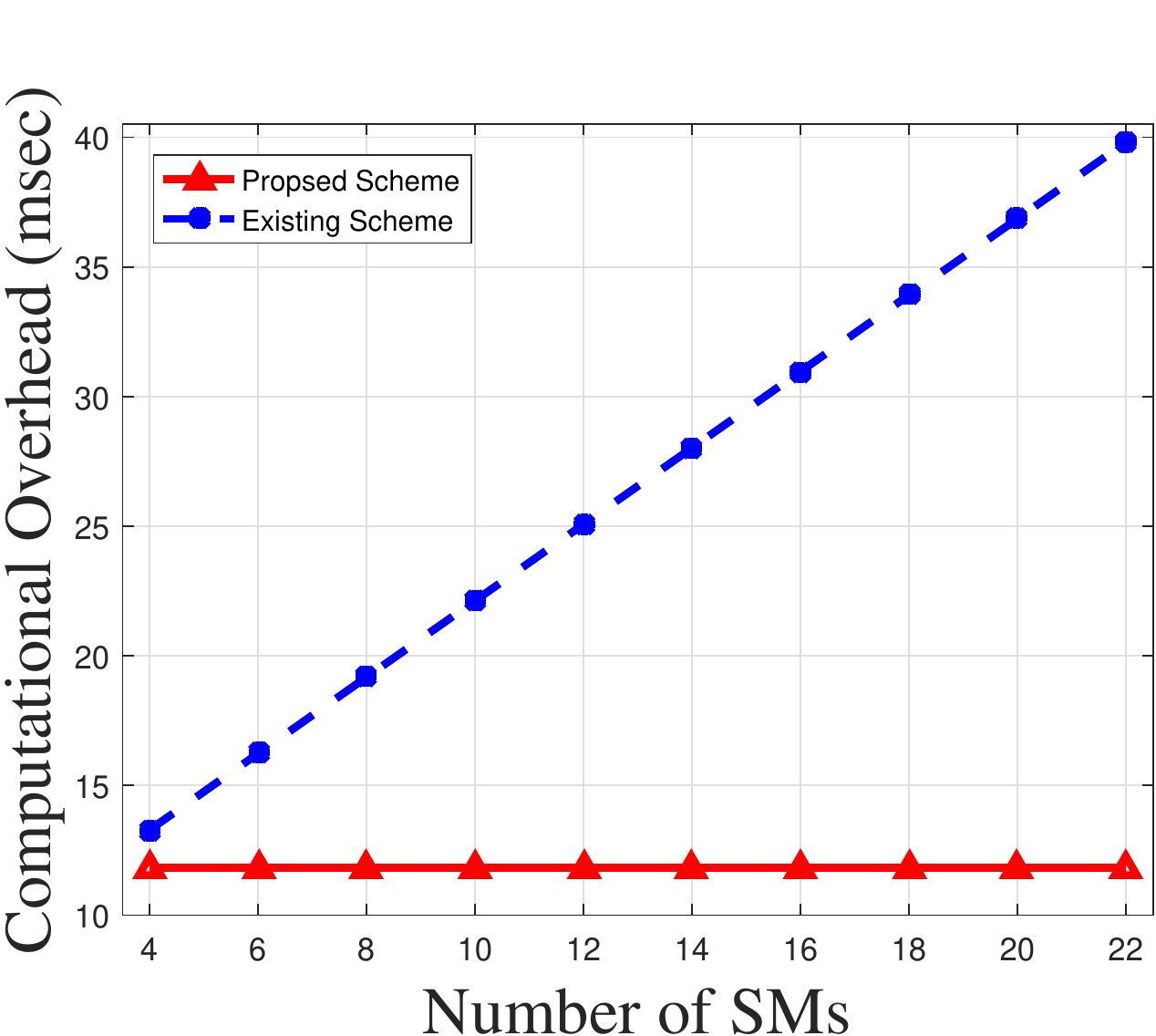}
		\caption{Computational overhead Vs number of SMs.}
		\label{fig:computationaloverheadsm}
	\end{subfigure}
	~
	\begin{subfigure}[t]{0.23\textwidth}
		\centering
		\includegraphics[scale=.33]{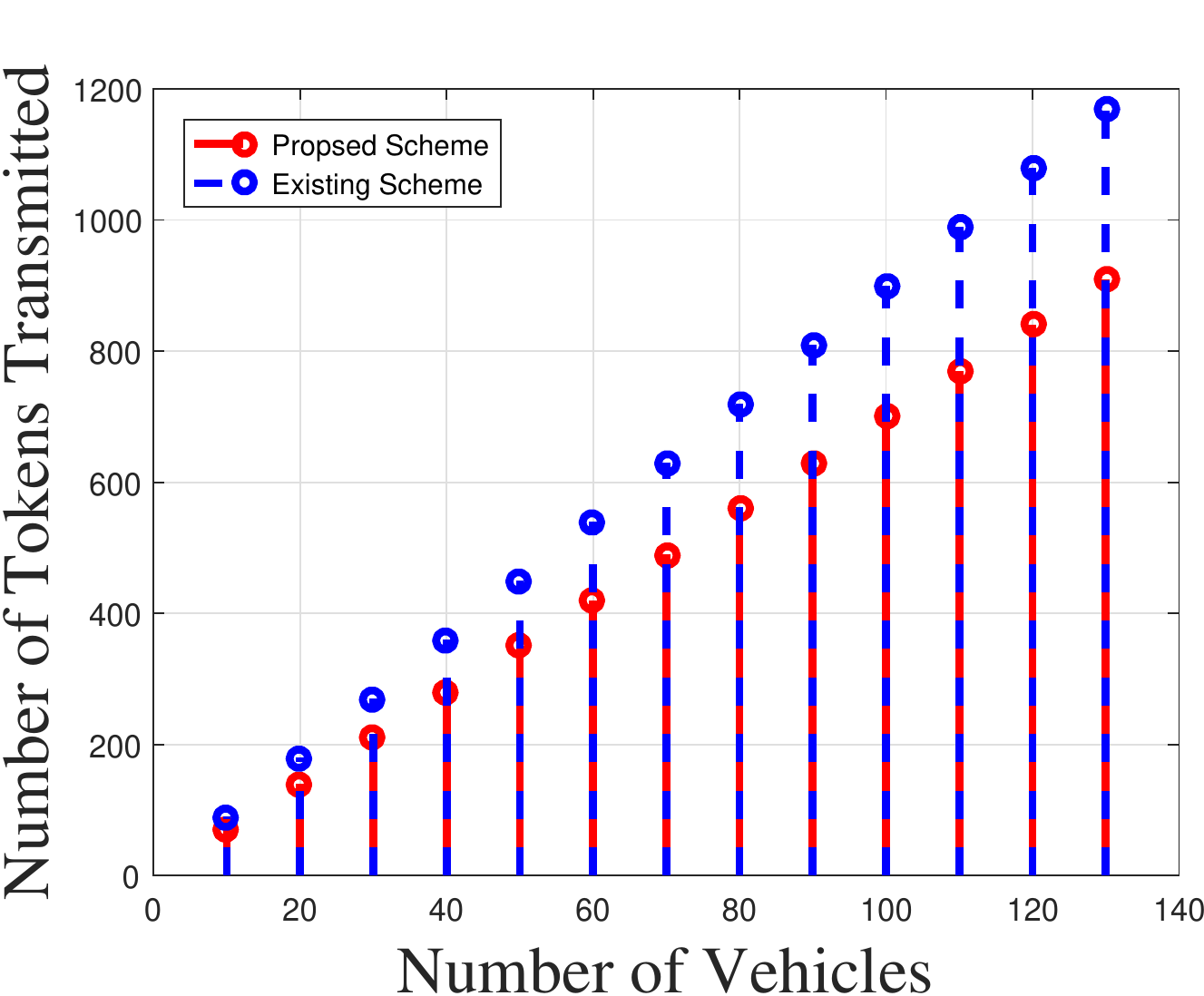}
		\caption{Communicational overhead Vs number of vehicles.}
		\label{fig:communicationaloverheadv}
	\end{subfigure}
	~
	\begin{subfigure}[t]{0.23\textwidth}
		\centering
		\includegraphics[scale=.33]{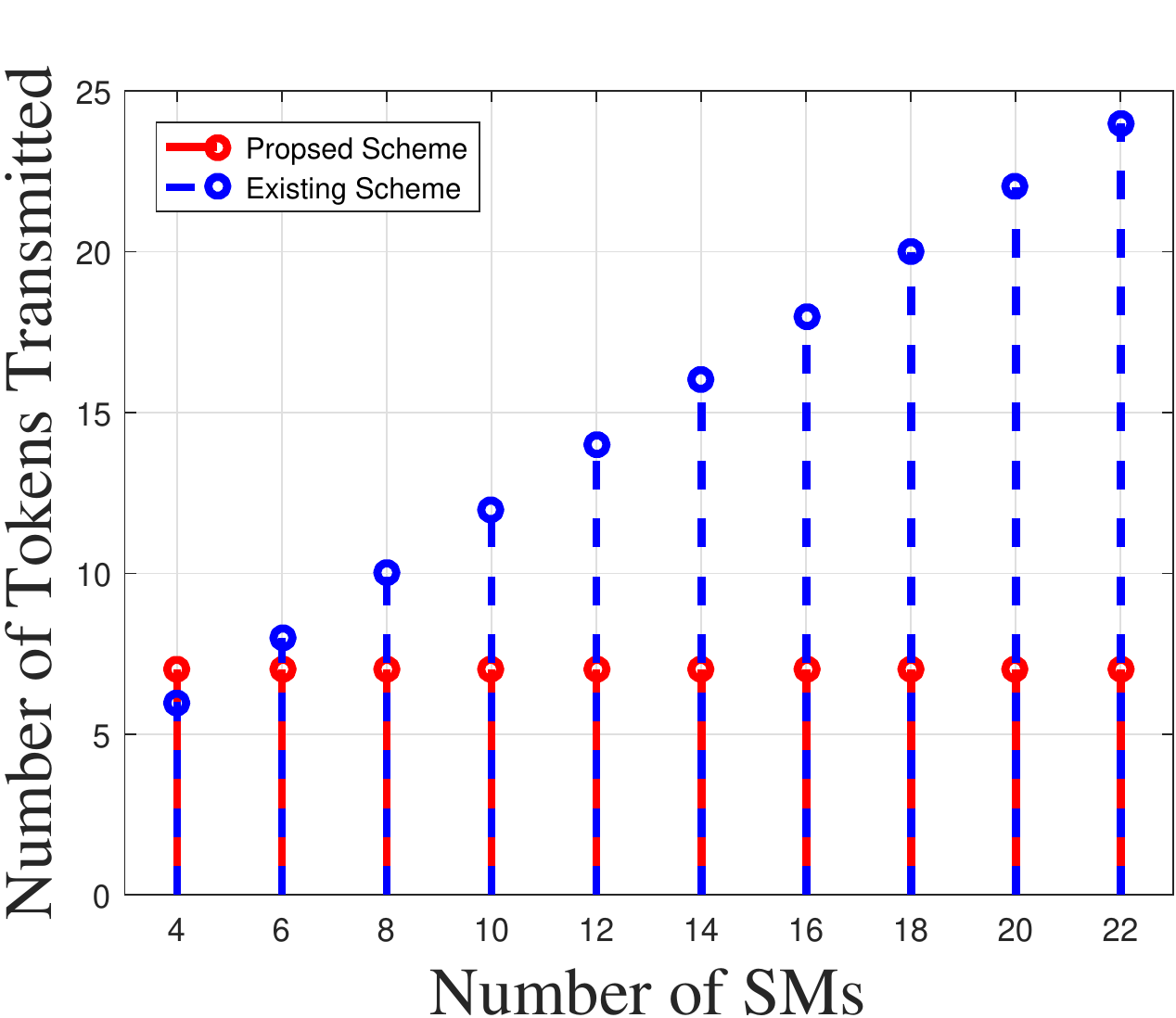}
		\caption{Communicational overhead Vs number of SMs.}		\label{fig:communicationaloverheadsm}
	\end{subfigure}
	\caption{An illustration of computational and communicational overhead analysis.}

\end{figure*}

In this section, the detailed analysis of the proposed scheme relative to an existing scheme \cite{8606185} has been carried out in terms of computational and communicational overheads analysis. These results have been validated typically for \textit{Phase III} since the majority of the computational and communicational overhead differences have been observed for this phase. In order to assess their performance, the experimental results derived by Yao \textit{et al.} in \cite{8606185} have been employed. The authors summarized the computational time required in executing the cryptographic hash function ($T_h$) and ECC-based multiplicative operation ($T_{ECC}$) in their work as 0.596 msec and 1.473 msec, respectively. The computational time with respect to XOR  and concatenation is assumed to be negligible. Further, the time for executing kdf operations has not been evaluated since it is used for session key generation that is performed only in the proposed scheme.\\
\indent In total, the proposed scheme executes a total of 5 hash functions and 6 ECC-based multiplicative operations, against a total of 5 and (k+3) operations respectively by the existing scheme. Thus, the total computational overhead associated with single pass of Phase III is $5 \times T_h + 6 \times T_{ECC}$ (in the proposed scheme) and  $5 \times T_h + (k+3) \times T_{ECC}$ (in the existing scheme). On the basis of this observation, Fig.~\ref{fig:computationaloverheadv} and \ref{fig:computationaloverheadsm} summarize the results with respect of variable number of vehicles and SMs. It is evident from Fig.~\ref{fig:computationaloverheadv} that the computational overhead increases when increasing the number of vehicles which can be attributed to the increasing number of authentication requests. However, the results clearly showcase that the proposed scheme depicts superior results  compared to the existing scheme. Additionally, the results obtained in Fig.~\ref{fig:computationaloverheadsm} clearly show that the proposed scheme is scalable and is not affected by the increase in the number  of SMs. On the other hand, the computational overhead  of the existing scheme rises significantly when increasing the number of SMs. \\
\indent Apart from the above comparison, Figs.~\ref{fig:communicationaloverheadv} and \ref{fig:communicationaloverheadsm} highlight the communicational overhead associated with the two schemes for a variable number of vehicles and SMs respectively. The comparison is based upon the number of token transmitted between the OBUs and SMs during the authentication process. For the proposed scheme, the following tokens $<\mathbb{A}uth_{OBU_i},~r_1,~T_{OBU_i}, ~\mathfrak{TD}_{OBU_i}, T_{SM_j}, ~r_2, ~\mathbb{A}uth_{SM_j}>$ were communicated between the two parties which results to a total of 7 tokens. Meanwhile, a total of $k+2$ tokens were sent in the existing scheme.
It is evident from the results that, for the scenarios, the existing scheme  experiences constant increase in the communicational overhead compared to the proposed scheme. It is worth noticing that the proposed scheme is unaffected by the number of SMs and and experiences less overhead even with increase in number of vehicles. Thus, it can be concluded that the proposed scheme is lightweight compared to the existing scheme.

\section{Conclusion} \label{sec:Conclusions}

VFC has emerged as a promising candidate for the provisioning of different services including safety and infotainment in modern ITS. However, its growing popularity is accompanied with a gigantic array of security attacks. Thus, it is essential to safeguard these networks against different attack vectors. In this direction, provably secure authentication and key exchange policies are essential. Therefore, in this paper, a lightweight authentication and key exchange mechanism for VFC infrastructures has been proposed. The designed mechanism is based on ECC and blockchain; which helps vehicles to seamlessly access VFS with the following features: cross-data center authentication, user anonymity, mutual authentication, lightweight, user privacy and confidentiality. The extensive performance assessment of the proposed mechanism established its superiority in terms of reduced communicational and computational overheads with enhanced security features relative to an existing scheme. 

\bibliographystyle{IEEEtran}
\bibliography{ICCRef1.bib}
\end{document}